\documentclass[11pt,letterpaper]{article}
\pdfoutput=1
\usepackage{amscd}
\usepackage{jheppub} 

\preprint{DAMTP-2018-08-14}

\title{Topological terms in Composite Higgs Models}

\author[a]{Joe Davighi}
\author[b]{and Ben Gripaios}

\affiliation[a]{Department of Applied Mathematics and Theoretical Physics, University of Cambridge, Wilberforce Road, Cambridge, UK}
\affiliation[b]{Cavendish Laboratory, University of Cambridge, J.~J.~Thomson Ave, Cambridge, UK}
\emailAdd{jed60@cam.ac.uk}
\emailAdd{gripaios@hep.phy.cam.ac.uk}

\abstract{
We apply a recent classification of topological action terms to Composite Higgs models based on a variety of coset spaces $G/H$ and discuss their phenomenology. The topological terms, which can all be obtained by integrating (possibly only locally-defined) differential forms, come in one of two types, with substantially differing  consequences for phenomenology. The first type of term (which appears in the minimal model based on $SO(5)/SO(4)$) is a field theory generalization of the Aharonov-Bohm phase in quantum mechanics. The phenomenological effects of such a term arise only at the non-perturbative level, and lead to $P$ and $CP$ violation in the Higgs sector. The second type of term (which appears in the model based on $SO(6)/SO(5)$) is a field theory generalization of the Dirac monopole in quantum mechanics and has physical effects even at the classical level. Perhaps most importantly, measuring the coefficient of such a term can allow one to probe the structure of the underlying microscopic theory. A particularly rich topological structure, with 6 distinct terms, is uncovered for the model based on $SO(6)/SO(4)$, containing 2 Higgs doublets and a singlet. Of the corresponding couplings, one is an integer and one is a phase.
}

\begin{document} 
\maketitle
\flushbottom

\section{Introduction}
The idea that the Higgs boson is composite remains an attractive solution to the electroweak hierarchy problem, albeit a slightly fine-tuned one. In the most plausible such models, the Higgs arises as a pseudo Nambu Goldstone boson (pNGB) associated with the breaking of an approximate global symmetry $G$ down to a subgroup $H$. Consequently, the Higgs mass would naturally reside somewhere below the energy scale associated with this symmetry breaking. Regardless of the details of the microscopic theory at high energies,
at low energies the presence of a mass gap separating the pNGBs from other, heavier resonances means that such other fields can be integrated out. The long distance physics is thus described by a non-linear sigma model on the homogeneous space $G/H$, parametrized by the pNGB fields. 

In order to successfully describe electroweak symmetry breaking, $G/H$ should satisfy the following requirements. Firstly, the linearly-realized subgroup $H$ should contain the electroweak gauge group, $SU(2)_L\times U(1)_Y$. Secondly, to guarantee consistency with electroweak precision measurements, specifically the mass ratio of the $W$ and $Z$ bosons, we shall require that $H$ contains the larger, custodial, symmetry $SU(2)_L \times SU(2)_R$ \cite{SIKIVIE1980189}, which is an accidental global symmetry of the Standard Model.\footnote{In fact, to prevent large corrections to the $Zb\bar{b}$ coupling, it is desirable to enlarge this even further \cite{Agashe:2006at}, though we will mostly ignore this nicety here.} Finally, in order to identify a subset of the pNGBs with the Composite Higgs, the spectrum of pNGBs parametrizing $G/H$ (which can be decomposed into irreducible representations of the unbroken symmetry group $H$) must contain at least one copy of the $(\mathbf{2},\mathbf{2})$ representation of the $SU(2)_L \times SU(2)_R$ subgroup.

Even after these requirements have been imposed, there remains a lengthy list of viable cosets with reasonable phenomenology; for example, $G/H=SO(5)/SO(4)$, $SO(6)/SO(5)$, and $SU(5)/SO(5)$ have all been explored extensively in the literature, due to various attractive features. A shortlist of candidates can be found, for example, in Table 1 of \cite{Mrazek:2011iu}. 

The physics of the Composite Higgs is directly analogous to that of pions, which are the degrees of freedom of QCD at long distances. Both theories are examples of four-dimensional sigma models on coset spaces $G/H$.
Recall that the QCD pions arise as the pNGBs associated with the spontaneously broken chiral symmetry ($G=SU(3)_L \times SU(3)_R$) of massless QCD down to its diagonal subgroup ($H=SU(3)_{\text{diag}}$), and thus live on the coset space $G/H=SU(3)_L \times SU(3)_R/SU(3)_{\text{diag}}\simeq SU(3)$. The action for the Composite Higgs is therefore to be constructed according to the same principles as the chiral lagrangian.
In both cases, the action consists {\em a priori} of all Lorentz invariant operators that can be constructed out of the pNGB fields, which are moreover invariant under the transitive $G$ action on $G/H$. The construction of Callan, Coleman, Wess and Zumino (CCWZ) \cite{Callan:1969sn} provides a systematic scheme for writing down such operators, arranged in order of increasing irrelevance at low energies, by using the additional structure of a $G$-invariant metric on $G/H$. However, this construction misses $G$-invariant terms in the action which are topological, in the sense that they require neither a metric on $G/H$ nor on spacetime.\footnote{As a result, such terms are invariant not just under the isometry group of spacetime, {\em viz.} the Poincar\'{e} group, but under the much larger group of orientation-preserving diffeomorphisms of spacetime.}

In the chiral lagrangian, the existence of such a topological term plays an essential role in pion physics. This topological term is the Wess-Zumino-Witten (WZW) term \cite{Wess:1971yu,Witten:1983tw}. It is constructed out of the sigma model fields $g(x)\in SU(3)$ from the $SU(3)_L \times SU(3)_R$-invariant closed 5-form,
$\omega = \frac{n}{240\pi^2}\mathrm{Tr}\ [(g^{-1}dg)^5]$, where $n\in\mathbb{Z}$.
While the action is not the integral of any local lagrangian over the 4-d spacetime $\Sigma$, it can nonetheless be written by integrating the 5-form $\omega$ over a 5-ball $B$ whose boundary is $\Sigma$. 

The WZW term is needed to reproduce the axial anomaly occurring in massless QCD with three flavours, which is not renormalized and so must be reproduced in the low energy effective theory (the chiral lagrangian). The precise matching of the anomaly coefficient constrains the integer coefficient $n$ of the WZW term to be equal to the number of colors in the UV gauge theory, in this case $n=N_C=3$. Upon gauging electromagnetism, the WZW term gives rise to, amongst other terms, a coupling of the neutral pion $\pi^0$ to $F\tilde{F}$, and so facilitates the axial current decay $\pi^0\rightarrow \gamma\gamma$, with the correct decay rate as measured by experiments. Turning this argument on its head, the form of the WZW term allows physicists to measure the number of colours in QCD, which is an integer, by measuring the rate of pion decay.

The WZW term performs a second crucial r\^{o}le in pion physics: it
is the leading order term in the chiral lagrangian that violates the discrete $\mathbb{Z}_2$ symmetry $(-1)^{N_B}$ which counts the number of pions modulo two. It therefore provides the dominant contribution to certain pion scattering processes which violate this discrete symmetry, for example $2\rightarrow 3$ decays such as $K^+ K^- \rightarrow \pi^0 \pi^+ \pi^-$, and decays of the $\Phi$ ($s\overline{s}$) meson to $K^0 \overline{K}^0$.

Given the important r\^{o}les played by the WZW term in the phenomenology of pions, it is natural to expect that topological terms may play an equally important r\^{o}le in the phenomenology of the Composite Higgs. Motivated by the r\^{o}le of the WZW term in the chiral lagrangian, as we have just discussed, we now describe more explicitly some possible ways in which topological terms could be important in the case of a Composite Higgs.

Firstly, it is worth pointing out that in the Standard Model, the Higgs lives on the flat, non-compact space $\mathbb{C}^2$. In contrast, a composite Higgs lives (typically) on a compact space $G/H$ (for example, a 4-sphere), which is only locally diffeomorphic to $\mathbb{C}^2$; topologically, $\mathbb{C}^2$  and (say) $S^4$ are very different beasts. Different coset spaces are distinguished from one another both by their local algebraic structure (which determines, for example, the representations in which the various pNGBs transform under the unbroken subalgebra $\mathfrak{h}$), but also by their differing global structures. Topological terms in the action allow us to probe these global properties of the Composite Higgs, which are intrinsically Beyond the Standard Model effects.

Just as we saw for the chiral lagrangian, the presence of a WZ-like term in the action (the definition of which we will make precise in \S \ref{review}) would yield unambiguous information about the UV theory from which the Composite Higgs emerges, via anomaly matching (which is not renormalized). To put this statement in a concrete setting, we first recall that certain Composite Higgs theories are favoured because they are believed to arise at low energies from gauge theories in the UV which contain only fermions ({\em i.e.} from theories which are free of fundamental scalars, and thus free of hierarchy problems of their own). For example, it appears that the $SO(6)/SO(5)$ model can be reached in the flow towards the IR from a gauge theory with gauge group $Sp(2N_C)$, for some number of colours $N_C$, with four Weyl fermions transforming in the fundamental representation of $Sp(2N_C)$. The argument for this is that this gauge theory has an $SU(4)\simeq SO(6)$ (where $\simeq$ here denotes local isomorphism) global flavour symmetry, corresponding to unitary rotations of the four fermions amongst themselves, which can be spontaneously broken to an $Sp(4)\simeq SO(5)$ subgroup by giving a vev to the fermion bilinear \cite{Barnard:2013zea}. 

Now, a gauge theory with a symplectic gauge group cannot suffer from a chiral anomaly, so by anomaly matching, the corresponding low energy Composite Higgs model should also be anomaly free. Now, as we shall see in \S \ref{6over5}, there is in fact a WZ term in the $SO(6)/SO(5)$ Composite Higgs theory, which can be written by integrating the $SO(6)$-invariant volume form on $S^5$ over a 5-dimensional submanifold whose boundary 
is the 4-dimensional worldvolume. Moreover, this WZ term reproduces the anomaly \cite{Gripaios:2016mmi}. Hence, we conclude that, if the $SO(6)/SO(5)$ Composite Higgs theory does indeed derive from a gauge theory with symplectic gauge group, then the WZ term must have its coefficient set to zero for consistency. Reversing the argument, if the WZ term in the low-energy $SO(6)/SO(5)$ sigma model were measured to be \textit{non}-zero, this would tell us that the UV completion cannot be the $Sp(2N_C)$ theory! Thus, we see that topological terms in the sigma model can provide us with pertinent probes of the UV theory. 

More generally, in any Composite Higgs model which has a viable UV completion in the form of a gauge theory (with only fermions), one must reproduce the chiral anomaly present (or not) in the gauge theory at low energies via a WZ term in the $G/H$ sigma model.

We now give an altogether different example which demonstrates the potential importance of topological terms to the Composite Higgs. In \cite{Gripaios:2016mmi}, the effect of a WZ term in a Composite Higgs model with the coset space $SO(5)\times U(1)/SO(4)$ was discussed. This model features a singlet pNGB, $\eta$, in addition to the complex doublet identified with the SM Higgs. The (gauged version of the) WZ term that was identified was found to dominate the decay of this singlet, as well as facilitating otherwise extremely rare decays such as $\eta \rightarrow h W^+ W^- Z$.\footnote{This model was originally proposed as a potential explanation for the resonance observed at 750 GeV in the diphoton channel, subsequently found to be but a statistical fluctuation.} In fact, as the present authors found in \cite{Davighi:2018inx}, the addition of this putative WZ term turns out to break the $U(1)$-invariance of the theory, and so there would in fact be no light $\eta$ boson at all if the WZ term were turned on. Nonetheless, it remains generally true that topological terms can provide the dominant decay channels for pNGBs in the low energy theory.

Although it is peripheral to the main thrust of this paper, it would be remiss of us not to remark that there may exist other topological effects in Composite Higgs models, albeit ones not directly associated to terms in the action. One such possible effect is the existence of topological defects analogous to the skyrmion, which plays the r\^{o}le of the baryon in the chiral lagrangian. If the third homotopy group of $G/H$ vanishes, then one expects there to exist topologically stable solutions to the classical equations of motion which correspond to homotopically non-trivial maps from a worldvolume with the topology $S^3\times S^1$ to $G/H$. This occurs, for example, in the ``littlest Higgs" theory based on the coset $SU(5)/SO(5)$, which has $\pi_3(SU(5)/SO(5))=\mathbb{Z}_2$. Being stable, the skyrmions have been suggested as a candidate for Dark Matter \cite{PhysRevD.80.074009,Gillioz:2010mr}. 

Given the possible physical effects, it is evidently useful to be able to find all possible topological terms in a given Composite Higgs model. In this paper, we shall try to answer this question in a more-or-less systematic fashion. In a recent paper \cite{Davighi:2018inx}, we suggested a homology-based classification of topological terms appearing in a generic sigma model (in any spacetime dimension) on an arbitrary homogeneous space $G/H$.
In this paper, we apply this formalism to classify the topological terms appearing in a selection of well-studied Composite Higgs cosets $G/H$. To wit: $SO(5)/SO(4)$, $SO(6)/SO(5)$, $SO(5)\times U(1)/SO(4)$, $SO(6)/SO(4)$, $SO(6)/SO(4)\times SO(2)$, and $SU(5)/SO(5)$. 
We find different results to those claimed earlier in the literature for four of these six models. Sometimes these differences are rather subtle from the phenomenological perspective, such as in the case of the Minimal Model (with coset $SO(5)/SO(4)$), while sometimes they are rather more drastic, such as in the case $SO(5)\times U(1)/SO(4)$. 
In the case of $SO(6)/SO(4)$, a rather rich topological structure is uncovered.

The structure of the paper is as follows. We begin in \S \ref{review} by reviewing the formalism developed in \cite{Davighi:2018inx}, and summarizing the main results which are relevant to the case of Composite Higgs models. We then tackle the cosets of interest one by one, in \S\S \ref{minimal}--\ref{littlest}. Each of the cosets chosen reveals its own distinct topological story. In \S \ref{connections}, we discuss how the different Composite Higgs models can be deformed into one another by the addition of explicit symmetry breaking operators; we show explicitly how, in one case, the topological terms identified in the different theories can be matched onto each other.
Finally, we conclude in \S \ref{discussion}.

\section{Review: Aharonov-Bohm and Wess-Zumino terms \label{review}}

We begin by summarizing the key results of \cite{Davighi:2018inx} that we need in the context of Composite Higgs models.

As we have already indicated, a Composite Higgs model is described by a sigma model, which is a quantum field theory whose degrees of freedom are maps $\phi$, in this case from a $4$-dimensional spacetime manifold $\Sigma^4$ into a coset space $G/H$ acted upon transitively by a Lie group $G$, which are the global symmetries of the theory. If we take $\Sigma^4$ to be an oriented, compact, connected manifold without boundary,\footnote{Compactness of $\Sigma^4$ can be justified as follows. When we Euclideanize the theory on $\mathbb{R}^4$, the usual leading-order, two-derivative kinetic term will force the fields to tend to a constant value `at infinity' in $\mathbb{R}^4$, in order that the action be finite. Thus, at least for the purposes of studying topological terms, we can one-point-compactify $\mathbb{R}^4$ to $S^4$.} then $\Sigma^4$ defines a class of $4$-cycles, in the sense of smooth singular homology, called its fundamental class $[\Sigma^4]$. One can then define a topological term in the sigma model action by integrating differential forms of degree $4$ on a $4$-cycle $z\in[\Sigma^4]$. Such $4$-forms can be readily supplied by pulling back $4$-forms from the target space using $\phi^*$. Completely equivalently, we can push forward the 4-cycle $z$ to a cycle $\phi_* z$ in $G/H$, on which we can directly integrate forms on $G/H$. We shall from hereon rename the cycle $\phi_* z$ in $G/H$ to be just $z$.\footnote{We remark that topological terms constructed using homology in this way can be defined on {\em all} compact, oriented, connected 4-manifolds without boundary, rather than just on $S^4$. This is necessary not only to be able to give a full description of physics (e.g. the dynamics in the background of a skyrmion requires us to consider $\Sigma = S^3 \times S^1$ \cite{Witten:1983tx}), but also to be able to couple to quantum gravity, in which the topology of spacetime may fluctuate. The construction is thus, in this sense at least, superior to the original constructions \cite{Finkelstein:1968hy,Witten:1983tw} based on homotopy.}

Subject to these assumptions, there are two types of topological terms, which we shall refer to as Aharonov-Bohm (AB) terms (also known in the literature as `theta terms') and Wess-Zumino (WZ) terms. Both types of terms shall play important r\^{o}les in Composite Higgs models. The essentials are as follows. 

An AB term is the integral over $z$ of a closed 4-form $A$, with a $U(1)$-valued coefficient:
\begin{equation}
S_{AB}[z]=\frac{\theta}{2\pi}\int_z A, \qquad dA=0, \quad \theta\in[0,2\pi). \label{ab term}
\end{equation}
In our normalization of choice, for which the action phase appearing in the path integral is $e^{2\pi i S[z]}$ (in other words, $h=1$), the 4-form $A$ is integral.\footnote{An integral $p$-form is one whose integral over any $p$-cycle is an integer. The normalization is such that, when $\theta$ is shifted by $2\pi$, the action $S_{AB}$ shifts by an integer for every cycle, leaving $e^{2\pi i S[z]}$ invariant; hence, $\theta$ and $\theta+2\pi$ are identified.}
Since an AB term is locally just a total derivative, it has no effect in the corresponding classical theory or in perturbation theory; thus its effects, if any, must be non-perturbative. In all these senses, an AB term is the sigma model analogue of the theta term in Yang-Mills theory.
The integral in (\ref{ab term}) only depends on the cohomology class of $A$ ({\em i.e.} it vanishes if $A$ is exact), hence AB terms exist only if the 4th de Rham cohomology of $G/H$ is non-vanishing. In general, the space of AB terms is classified by the quotient $H_{dR}^4(M,\mathbb{R})/H_{dR}^4(M,\mathbb{Z})$. In particular, this means there is a topological term in the minimal model whose target space is $SO(5)/SO(4)\simeq S^4$. We shall discuss in detail the phenomenological effects of this AB term in \S \ref{minimal}.

A WZ term is more subtle, and involves integrating 4-forms on $G/H$ that are not closed, and which may be only locally-defined. If the worldvolume 4-cycle $z$ is a boundary $z=\partial b$, the action may  be written straightforwardly as the integral of a $G$-invariant, integral,\footnote{This integrality requirement for WZ terms is to ensure the action phase $e^{2\pi i S[z]}$, for a given worldvolume cycle $z$, is free of ambiguities. If $z$ is a boundary, this corresponds to ambiguities in the possible choices of 5-chain $b$ such that $z=\partial b$.} closed 5-form $\omega$ over the 5-chain $b$. This is not possible for cycles which are not boundaries; in such cases, the appropriate language for formulating the WZ action in terms of local integrals is \v Cech cohomology.\footnote{Alternatively, and more elegantly, the action phase $e^{2\pi i S[z]}$ for a WZ term is 
a differential character, as defined by Cheeger-Simons \cite{10.1007/BFb0075216}, of which the $(p+1)$-form $\omega$ is the curvature. Note that, in this language, an AB term is also a differential character, but one for which the curvature is zero.}
For the details of this construction, we refer the reader to \cite{Davighi:2018inx} (see also \cite{Alvarez:1984es}). For our purposes here, however, we nevertheless think it important to highlight two facts about WZ terms which are not widely appreciated in the literature.  

Firstly, and contrary to what one may read in the literature, the existence of WZ terms in a $p$-dimensional sigma model does not require the $(p+1)$th de Rham cohomology be non-vanishing. Indeed, one can readily see this from the simple example of quantum mechanics of a point particle on the plane, which can be formulated as a sigma model with worldvolume dimension $p=1$, and target space $G/H=\mathbb{R}^2$. In this example, there is a WZ term corresponding to the closed, translation-invariant 2-form $F=Bdx\wedge dy$, which is of course exact because the cohomology of $\mathbb{R}^n$ is trivial. The 2-form $F$ may here be identified with the electromagnetic field strength, for a constant magnetic field out of the plane. The addition of this topological term to the action modifies the spectrum of the Hamiltonian from that of a free particle, to the Landau level spectrum. Thus, there is a WZ term, with profound physical effects on both the classical and quantum theories, despite $H^2_{dR}(G/H)=0$; it corresponds to the existence of a closed, $G$-invariant 2-form on $G/H$.
Returning to the case of our 4-d Composite Higgs model, WZ terms \textit{may} exist for every $G$-invariant, integral, closed 5-form $\omega$ on $G/H$. This 5-form may be exact (in which case the coefficient will be $\mathbb{R}$-valued, since every exact form is automatically an integral form), or not exact (in which case the coefficient will be $\mathbb{Z}$-valued for the form to be integral). 

The second fact is that mere $G$-invariance of the 5-form is not sufficient to guarantee the existence of a WZ term in the action which is $G$-invariant. As we shall see in \S \ref{anomaly}, the $G$ symmetry can be anomalous for topological reasons in the case where $H_4(G/H,\mathbb{Z})\neq 0$, that is, if there are non-trivial 4-cycles in $G/H$ corresponding to the existence of physical worldvolumes that are not boundaries. Rather, in general, it was shown in \cite{Davighi:2018inx} that the necessary \textit{and} sufficient condition for a $G$-invariant WZ term (at least when $G$ is connected) is that 
\begin{equation}
\iota_X \omega = df_X, \quad f_X\in \Lambda^{3}(G/H),\ \forall X\in \mathfrak{g},
\end{equation}
where $\Lambda^{3}(G/H)$ denotes the space of 3-forms on $G/H$, and the set of vector fields $\{X\}$ are the generators 
of the $G$ action on $G/H$.\footnote{Each vector field $X$ is the pushforward to $G/H$ of a right-invariant vector field on $G$, under the canonical projection map $\pi:G\rightarrow G/H$.} We refer to this condition, as we did in \cite{Davighi:2018inx}, as the Manton condition.\footnote{Provided the Manton condition is satisfied, the 3-forms $\{f_X\}$ define, via their integrals over appropriate spatial hypersurfaces, the contributions to the Noether charges for $G$-invariance from the WZ term \cite{Davighi:2018inx}.}
Moreover, it was shown in \cite{Davighi:2018inx} that the Manton condition turns out to be necessarily satisfied for all $X\in[\mathfrak{g},\mathfrak{g}]$; in particular, the Manton condition is satisfied for all of $\mathfrak{g}$ when $G$ is a semi-simple Lie group.

This useful result tells us that left-invariance of the closed 5-form $\omega$ is sufficient to guarantee the existence of an invariant WZ term in all but one of the Composite Higgs models that follow (even in the presence of homologically non-trivial 4-cycles, as in \S \ref{6over4}). The exception is the model considered in \S \ref{anomaly}, in which the group $G=SO(5)\times U(1)$ is not semi-simple; indeed, in this case, we find that invariance under the $U(1)$ factor of $G$ is broken (in the quantum theory) by the addition of the WZ term. 

We now turn to classifying the topological terms appearing in our list of phenomenologically relevant Composite Higgs models, in (approximate) order of increasing difficulty. While we briefly pointed out in \cite{Davighi:2018inx} the examples that we shall discuss in \S\S \ref{minimal}, \ref{6over5}, and \ref{anomaly}, in this paper we wish to study them (and others) more comprehensively.
We begin with the minimal model.

\section{The Aharonov-Bohm term in the $SO(5)/SO(4)$ model \label{minimal}}

The minimal Composite Higgs model (MCHM) \cite{Agashe:2004rs} is a sigma model whose target space is
$G/H=SO(5)/SO(4)\simeq S^4$. 
There are no non-trivial 5-forms on the target, it being a 4-manifold, and so there are no WZ terms in the minimal model.
However, since $H_{dR}^4(S^4,\mathbb{R})=\mathbb{R}$, and $H_{dR}^4(S^4,\mathbb{Z})=\mathbb{Z}$, there is an AB term given by the integral of a 4-form proportional to the volume form on $S^4$. 

In terms of the Higgs doublet fields $H=(h_1,h_2,h_3,h_4)$ which transform in the fundamental representation of the linearly-realized $SO(4)$ subgroup, and which provide local coordinates on the $S^4$ target space ({\em i.e.} coordinates only a patch of $S^4$, albeit a rather large patch which covers all but a finite set of points), the contribution to the AB term from a local patch may be written
\begin{equation}
S_{AB}=\frac{\theta}{2\pi}\int \frac{1}{V_4}dh_1\wedge dh_2\wedge dh_3\wedge dh_4, \quad \theta \in [0,2\pi), \label{AB in MCHM}
\end{equation}
where $V_4$ is the volume of the unit 4-sphere, and $dh_1\wedge dh_2\wedge dh_3\wedge dh_4$ denotes the volume form on $S^4$.\footnote{For the reader who seeks an explicit expression for the lagrangian in this example, one may of course pull-back the 4-form in (\ref{AB in MCHM}) to obtain the $SO(5)$-invariant lagrangian density
$$\mathcal{L}_{AB}\left(h_i(x^\mu)\right)=\frac{\theta}{2\pi}\epsilon^{\mu\nu\rho\sigma}\partial_\mu h_1 \partial_\nu h_2 \partial_\rho h_3 \partial_\sigma h_4,$$ such that $S_{AB}=\int d^4 x\ \mathcal{L}_{AB}$, where $x^\mu$ are coordinates on the worldvolume. Of course, $\mathcal{L}_{AB}$ is locally a total derivative as for any AB term.
}
The space of inequivalent topological action phases is thus
$\mathbb{R}/\mathbb{Z}=U(1)$, labelled by the coefficient $\theta \sim \theta+2\pi$. The existence of a topological term in the MCHM, which we pointed out in \cite{Davighi:2018inx}, had gone previously unnoticed in the literature.

The effects of this term, like all AB terms, are entirely quantum-mechanical and non-perturbative. Unlike the theta term in 2-d sigma models, whose physical effects are largest in the deep infrared, we expect the effects of an AB term in a 4-d sigma model such as a Composite Higgs theory to become large in the ultraviolet. This conclusion follows from an instanton argument, which we summarize in Appendix \ref{instanton}. This raises an exciting prospect for searches at the TeV scale and beyond. However, by that same argument, at low energies the non-perturbative effects of this AB term in the MCHM are exponentially suppressed. Thus, whether there are any measurable effects at the energy scales probed by the LHC, say, is unclear.

Some hope in this direction comes from the fact that, as we will now show, the  AB term in the MCHM violates both $P$ and $CP$.\footnote{To be clear, we are not suggesting this AB term is the {\em leading order} term in the effective field theory expansion that breaks $P$ and $CP$, which it is certainly not: indeed, 4-derivative (non-topological) operators exist in the ordinary CCWZ construction which break these discrete symmetries. Note that in the effective field theory expansion of the sigma model action, the AB term may be regarded as an ``infinite order'' contribution, since it corresponds locally to a total derivative in the lagrangian.
} Violation of these symmetries in the Higgs sector is known to lead to effects in a variety of physical processes and is strongly constrained. Thus, even though the effects of the topological term at lower energies are expected to be small, they may, nevertheless, have observable consequences. If the angle $\theta$ in (\ref{AB in MCHM}) could be measured to be neither zero nor $\pi$, perhaps by observing some instanton-induced effect, then one would deduce that the microscopic theory the sigma model originates from breaks $P$ and $CP$.

\subsection{$P$ and $CP$ violation}

To see that $P$ and $CP$ are violated, we must first discuss how they are implemented in the $SO(5)/SO(4)$ model. The leading-order (two-derivative) term in the low-energy effective theory is built using the CCWZ construction and requires a metric on both the target space and the worldvolume. The metric on the target space $S^4$ should be invariant under the action of at least the group $G=SO(5)$, but such a metric (which is, of course, just the round metric on $S^4$) is, in fact, invariant under the full orthogonal group $O(5)$. Moreover, since this a maximal isometry group of 4-d manifolds, there is no larger group that can act isometrically. The metric on the worldvolume $S^4$ is just the Euclideanized version of the Minknowski metric on $\mathbb{R}^4$, which is also the round metric on $S^4$, itself with isometry group $O(5)$. The full symmetry of the two-derivative term is thus $O(5)\times O(5)$.

The usual parity transformation $P$ corresponds (in the Euclideanized theory) to the factor group $O(5)/SO(5) \simeq \mathbb{Z}/2\mathbb{Z}$ acting on the worldvolume. This is an orientation-{\em reversing} diffeomorphism of the worldvolume, and so the topological term,
which is proportional to the volume form on the worldvolume after pullback, 
changes sign under the action of $P$. It is invariant only for forms whose integral over $S^4$ is equal to $1/2$ (mod an integer), corresponding to $\theta=\pi$.

As for charge conjugation, it is defined in the Standard Model as the automorphism of the Lie algebra $\mathfrak{su(3)} \oplus \mathfrak{su}(2) \oplus \mathfrak{u}(1)$ corresponding to complex conjugation of the underlying unitary transformations that define the group and its algebra. We wish to extend this transformation to the composite sector in such a way as to obtain a $C$-invariant 2-derivative term. To do so, we may focus our attention on the electroweak subalgebra $\mathfrak{su}(2) \oplus \mathfrak{u}(1)$, which is embedded in the composite sector as a subalgebra of $\mathfrak{so}(4) \simeq \mathfrak{su}(2) \oplus \mathfrak{su}(2)$, corresponding to the algebra of $H = SO(4)$. Now, the automorphism of $\mathfrak{su}(2) \oplus \mathfrak{u}(1)$ corresponding to complex conjugation can be extended to an automorphism of $\mathfrak{su}(2) \oplus \mathfrak{su}(2)$, given explicitly by conjugating each $\mathfrak{su}(2)$ factor by the Pauli matrix $\sigma_2 := \left(\begin{smallmatrix} 0 & -i \\ i & 0 \end{smallmatrix}\right) = -ie^{i\frac{\pi}{2}\sigma_2}$. Neither of these automorphisms are inner (because $\mathfrak{u}(1)$ has no non-trivial inner automorphisms and because $\sigma_2 \notin SU(2)$), but the latter does induce an inner automorphism on the factor group $SO(4) \simeq (SU(2)\times SU(2))/(\mathbb{Z}/2 \mathbb{Z})$: it sends $SU(2)\times SU(2) \ni (a,b) \mapsto (\sigma_2 a \sigma_2^{-1}, \sigma_2 b \sigma_2^{-1}) \sim (-\sigma_2 a \sigma_2^{-1}, -\sigma_2 b \sigma_2^{-1}) =  (i\sigma_2 a i\sigma_2^{-1}, i\sigma_2 b i\sigma_2^{-1})$ (where $\sim$ denotes the $\mathbb{Z}/2 \mathbb{Z}$ equivalence). Hence the action on the factor group is equivalent to conjugation by $[(i \sigma_2, i\sigma_2)] \in (SU(2)\times SU(2))/(\mathbb{Z}/2 \mathbb{Z})$.  

Now, quite generally, an inner automorphism of $H$ by $h\in H$ defines an inner automorphism of $G \supset H$ as $G \ni g \mapsto hgh^{-1}$,  whose action on cosets, $G/H \ni gH\mapsto hgh^{-1}H = hgH$, is not only well-defined, but also is equivalent to the original action of $H \subset G$ induced by left multiplication in $G$ that is central to the discussion in this paper. Thus we see that we can not only naturally extend the definition of $C$ in the Standard Model to the MCHM (in a way that the leading order action term is manifestly invariant, even after we gauge the SM subgroup), but that doing so is equivalent to an action on $G/H$ by an element in $SO(4) \subset SO(5)$. 
Since the topological term is $SO(5)$-invariant by construction, it is invariant under $C$. Hence it changes by a sign under $CP$, except for forms whose integral over $S^4$ is equal to $1/2$ (mod an integer).

We remark that, just as for the parity transformation, the topological term also changes by a sign under the action of the factor group $O(5)/SO(5) \simeq \mathbb{Z}/2\mathbb{Z}$ on the target space. This symmetry has been exploited in the literature \cite{Agashe:2006at} to prevent unobserved corrections to the decay rate of the $Z$-boson to $b$-quarks, compared to the Standard Model prediction. We can see that it is incompatible with a non-vanishing topological term, except for forms whose integral over $S^4$ is equal to $1/2$ (mod an integer).

The physics associated with AB terms appearing in other Composite Higgs models follows a similar story to that discussed here in the context of the minimal model. To summarize, the essential features are (i) that AB terms are likely to violate discrete symmetries, such as $P$ and $CP$, and (ii) 
they can only affect physics at the non-perturbative level.

\section{The Wess-Zumino term in the $SO(6)/SO(5)$ model \label{6over5}}

Consider the Composite Higgs model based on the homogeneous space $G/H=SO(6)/SO(5)\simeq S^5$ \cite{Gripaios:2009pe}. The five pNGBs transforms in the fundamental representation of the unbroken $SO(5)$ symmetry, which decomposes under $SU(2)_L\times SU(2)_R$ as $(\mathbf{2},\mathbf{2})\oplus (\mathbf{1},\mathbf{1})$. Thus, in addition to the Higgs doublet $H=(h_1,h_2,h_3,h_4)$, there is a Standard Model singlet $\eta$ in this theory. The fields $(\eta,H)$ provide (local) coordinates on the $S^5$ target space.

The principal appeal of this model, compared to the minimal model, is that one can easily imagine a UV completion in the form of a (technically natural) strongly coupled $Sp(2N_c)$ gauge theory with four Weyl fermions transforming in the fundamental of the gauge group, which has $SU(4)$ flavour symmetry. An explicit realization of the necessary spontaneous symmetry breaking of $SU(4)$ down to an $Sp(4)\simeq SO(5)$ subgroup 
has been proposed in \cite{Barnard:2013zea}. An explicit formulation of the microscopic theory such as this would of course provide a unique prediction for the quantized coefficient of the WZ term in the $SO(6)/SO(5)$ Composite Higgs model, via anomaly matching.

The WZ term in this theory corresponds to the closed, integral, $SO(6)$-invariant 5-form $\omega$ on $S^5$, which is simply the volume form, as originally described in \cite{Gripaios:2009pe}. Indeed, a straightforward calculation using the relative Lie algebra cohomology cochain complex\footnote{We shall expand on the role of Lie algebra cohomology in \S \ref{WZ}.} reveals that this is the unique $SO(6)$-invariant 5-form on $SO(6)/SO(5)$, up to normalization (in fact, the volume form is the only $SO(6)$-invariant differential form on $S^5$ of \textit{any} positive degree). Thus, there is a single WZ term in this model.

The Manton condition is satisfied trivially here, because the fourth de Rham cohomology of $S^5$ vanishes, so the closed 4-forms $\iota_X \omega$ are necessarily exact. For the same reason, there are no AB terms. Since the fourth singular homology vanishes, we can always follow Witten's construction and write the action as the (manifestly $SO(6)$-invariant) integral of $\omega$ over a 5-ball $B$ whose boundary $z=\partial B$ is our worldvolume cycle:
\begin{equation}
S_{WZ}[z=\partial B]= \frac{n}{V_5} \int_B d\eta\wedge d^4 H,\quad n\in\mathbb{Z},
\end{equation}
where $d\eta \wedge d^4 H$ is short-hand for the volume form on $S^5$ in our local ``Higgs" coordinates $(\eta,H)$, with $d^4 H\equiv dh_1\wedge dh_2\wedge dh_3\wedge dh_4$, and $V_5=\pi^3$ is just the volume of a unit 5-sphere. As noted above, depending on the details of the microscopic theory, the integer coefficient $n$ will be fixed by anomaly matching.

What phenomenological effects are associated with this WZ term? Na\"ively, the WZ term is a dimension-9 operator, as can be seen by considering the action locally. The Poincar\'e lemma means we can write $\omega=dA$ on a local patch, for example
\begin{equation}
S_{WZ}[z]=\frac{n}{V_5} \int_z \eta\ dh_1\wedge dh_2\wedge dh_3\wedge dh_4,
\end{equation}
which contains 5 fields and 4 derivatives, and is thus dimension-9. We might therefore expect this operator to be entirely irrelevant to the phenomenology at low energies. However, in order to study the phenomenology, it is necessary to first gauge the Standard Model subgroup $SU(2)_L\times U(1)_Y \subset SO(5)$.

Gauging the WZ term is a subtle issue, because the 4-dimensional lagrangian for the WZ term (which, remember, is only valid in a local patch) is not $G$-invariant, but shifts by an exact form. This means that a na\"ive ``covariantization" of the derivative $d\rightarrow d-A$ does not yield a gauge-invariant action. 
The gauging of topological terms is a subtle problem, even in cases where the construction of Witten can be carried out \cite{Witten:1983tw,Hull:1989jk,HULL1991379,Chu:1996fr,PhysRevD.30.594}. We postpone the discussion the gauging of topological terms in the general case to future work, remarking here only that upon gauging, one expects the WZ term to give rise to operators of dimension-5 which couple the Composite Higgs fields to the electroweak gauge bosons $W^{\pm}$ and $Z$,\footnote{This is precisely analogous to the gauging of electromagnetism in the chiral lagrangian, which we discussed in the Introduction, which leads to the dimension-5 operator $\pi F \tilde{F}$ and thus pion decay to two photons.} which are certainly important to the TeV scale physics of this theory.

We now turn to a more subtle example, where the subtlety is concerning $G$-invariance of the putative WZ term.

\section{The $SO(5)\times U(1)/SO(4)$ model \label{anomaly}} 

Consider the Composite Higgs model on the coset space $G/H=(SO(5)\times U(1))/ SO(4)\simeq S^4\times S^1$, in which a WZ term was incorrectly identified \cite{Gripaios:2016mmi}. The error was that a WZ term was postulated due to the existence of a $G$-invariant 5-form, when it turns out that one cannot write down a corresponding $G$-invariant action (phase) for worldvolumes corresponding to homologically non-trivial 4-cycles. This was observed in \cite{Davighi:2018inx}, and we shall elaborate on the discussion in what follows.

The target space is homeomorphic to $S^4\times S^1$, which has non-vanishing 4th and 5th cohomology, so there are potentially both AB and WZ terms.\footnote{As we emphasized in the Introduction, WZ terms are strictly in correspondence with invariant 5-cocycles, and not de Rham cohomology classes; nonetheless, because $G/H$ is compact and $G$ is connected, the non-vanishing fifth de Rham cohomology of $G/H$ implies that there is a $G$-invariant 5-form on $G/H$ \cite{0031.24803}.} The potential problem with $G$-invariance of the putative WZ term arises due to the non-trivial 4-cycles in $G/H$ which wrap around the $S^4$ factor, which mean that Witten's construction cannot be applied; moreover, the group $G$ is not semi-simple because of the $U(1)$ factor. This means we will have to check the Manton condition explicitly. Indeed, the $SO(5)\times U(1)$-invariant, closed, integral 5-form $\omega$, which is just the volume form on $S^4\times S^1$, fails to satisfy the Manton condition for the generator of $U(1)\subset G$,\footnote{The interior product of the volume form on $S^4\times S^1$ with the vector field generating the $U(1)$ factor is proportional to the volume form on the $S^4$ factor, which is closed but not exact.} and so the putative WZ term in fact explicitly breaks $U(1)$-invariance. Thus, there is no such WZ term.
\footnote{
We would like to emphasize that there is nonetheless a WZ term in the corresponding classical theory. The classical equations of motion, obtained by variation of the action, only depend on the WZ term through the 5-form $\omega$; thus, classical $G$-invariance is implied by $G$-invariance of $\omega$.
It is only in the quantum theory that we require the action itself (or, more precisely, the action phase $e^{2\pi i S[z]}$) to be $G$-invariant, and this requires the stronger Manton condition be satisfied. The difference between $G$-invariance of $\omega$ ({\em i.e.} $L_X\omega = 0$) and the Manton condition ({\em i.e.} $\iota_X \omega = df_X$) is purely topological, depending only on global information. The quantum theory is sensitive to this global information,  whereas the classical theory is not.
}

To see more explicitly how the problem with $U(1)$ invariance arises, we again introduce local Higgs coordinates $(\eta,H)$, where now $\eta\in S^1$, and the Higgs field provides local coordinates on the $S^4$ factor. Consider a worldvolume which corresponds to a non-trivial 4-cycle $z$ in the target space; for example, let $z$ wrap the $S^4$ factor some $W$ times, at some fixed value of the $S^1$ coordinate, $\eta_0$. On this cycle, we may write $\omega=dA$, where $A\propto \eta_0 d^4 H$ is well-defined on $z$ (again, $d^4H$ is shorthand for the volume form on the $S^4$ factor), and the WZ term is then given by the integral
\begin{equation}
\frac{n}{2\pi V_{4}}\int_z \eta_0\ d^4 H = \frac{n}{2\pi}\eta_0 W, \label{integral}
\end{equation}
where $V_{4}=\frac{8}{3}\pi^2$ is the volume of the 4-sphere (the factor $2\pi V_{4}$ is just the volume of the target space, such that $n\in\mathbb{Z}$ corresponds to $\omega$ being an integral form). This is clearly not invariant under the action of $U(1)$ on this cycle, which shifts $\eta_0 \rightarrow \eta_0 + a$ for some $a\in [0,2\pi)$. However, the $U(1)$ symmetry is not completely broken, because the action phase $e^{2\pi i S[z]}$ remains invariant under discrete shifts (for any $W$), such that $an\in 2\pi\mathbb{Z}$. Thus, the symmetry of the corresponding classical theory is broken, due to the WZ term, from 
\begin{equation}
SO(5)\times U(1)\rightarrow SO(5)\times \mathbb{Z}/n\mathbb{Z}
\end{equation}
in the quantum theory. This is directly analogous to the breaking of translation invariance that occurs (in the quantum theory) upon coupling a particle on the 2-torus to 
a translation invariant magnetic field, a fact which was first observed by Manton \cite{Manton:1985jm}. This instructive example from quantum mechanics, and its connection to this Composite Higgs model, was discussed in \cite{Davighi:2018inx}.

There is nonetheless still an AB term in this model, equal to $(\theta/2\pi)\int_z \frac{1}{V_4}d^4 H$, where $\theta\sim \theta+2\pi$, which counts the winding number into the $S^4$ factor of the target.

\section{The $SO(6)/SO(4)$ model \label{6over4}}

In this section, we turn to a model with a very rich topological structure, based on the coset $SO(6)/SO(4)$. As we shall soon see, this model exhibits both AB and WZ terms, in a non-trivial way.

The spectrum features two Higgs doublets, in addition to a singlet $\eta$. This model is attractive partly because the coset space is isomorphic to $SU(4)/SO(4)$, and this global symmetry breaking pattern may therefore be exhibited by an $SO(N_c)$ gauge theory with 4 fundamental Weyl fermions. A closely related model was discussed at length in \cite{Mrazek:2011iu}, which quotients by a further $SO(2)$ factor, thus removing the additional scalar. We will turn to that model in \S \ref{6over2times4}.

From our topological viewpoint, the manifolds $SO(n)/SO(n-2)$\footnote{The manifold $SO(n)/SO(n-2)$ is an example of a Stiefel manifold. It is the space of orthonormal $2$-frames in $\mathbb{R}^n$.} are rather unusual, in that, for even $n$, they have two non-vanishing cohomology groups, in neighbouring degrees $n-2$ and $n-1$.
This occurs, somewhat serendipitously, at the 4th and 5th cohomologies when $n=6$, which is the particular case of interest as a Composite Higgs model for group theoretic reasons.\footnote{There is a low-dimensional analogue of this problem, which is a $p=2$ sigma model ({\em i.e.} describing a string) with target space $SO(4)/SO(2)$, for which non-vanishing $H_{dR}^2$ yields an AB term, and non-vanishing $H_{dR}^3$ implies there is at least one WZ term. This model may be studied as a useful warm-up for the Composite Higgs example discussed in the text!}

In order to elucidate the topological structure of this theory, it is helpful to first describe the geometry of this target space. For any integer $n\geq 3$, the homogeneous space $SO(n)/SO(n-2)$ can be realised as a fibre bundle over $S^{n-1}$ with fibre $S^{n-2}$, namely the unit tangent bundle of $S^{n-1}$, which can be described by a point on $S^{n-1}$ and a unit tangent vector at that point. To see this, observe that $SO(n)$ has a transitive action on this space (induced by the usual action on $\mathbb{R}^n$), with stabilizer $SO(n-2)$. Indeed, the point on $S^{n-1}$ is stabilized by $SO(n-1)$, while a given unit vector tangent to that point gets moved by $SO(n-1)$, but is stabilized by the subgroup $SO(n-2)\subset SO(n-1)$. Thus, by the orbit-stabilizer theorem, the unit tangent bundle is isomorphic to the homogeneous space $SO(n)/SO(n-2)$.

Our target space $SO(6)/SO(4)$ is thus a 4-sphere fibred over a 5-sphere, and it is helpful to define the projection map for this bundle (which we shall on occasion refer to as $E$ for brevity):
\begin{equation}
\pi: E\equiv SO(6)/SO(4)\rightarrow S^5,
\end{equation}
with which we can pull-back ($\pi^*$) forms from $S^5$ to $E$, and also push-forward ($\pi_*$) cycles in $E$ to cycles in the base $S^5$. 
The non-vanishing homology groups 
\begin{equation}
H_{4}(E,\mathbb{Z})=H_{5}(E,\mathbb{Z})=\mathbb{Z} \label{homology}
\end{equation}
are generated by cycles which wrap the $S^4$ fibre and the $S^5$ base respectively.\footnote{These claims may be proven by considering the Gysin and Wang exact sequences in homology for the bundle $S^4\rightarrow E \rightarrow S^5$. The Gysin sequence is
$$ \dots \rightarrow H_1(S^5) \rightarrow H_5(E) \xrightarrow{\pi_*} H_5(S^5) \rightarrow H_0(S^5) \rightarrow H_4 (E) \xrightarrow{\pi_*} H_4(S^5)\rightarrow \dots,$$ where $\pi$ denotes the bundle projection, which reduces to
$$ 0 \rightarrow \mathbb{Z} \xrightarrow{i=\pi_*} \mathbb{Z} \xrightarrow{j} \mathbb{Z} \xrightarrow{k} \mathbb{Z} \xrightarrow{\pi_*} 0.$$ From the fact that this is an exact sequence, we can deduce that the map $\mathbb{Z} \xrightarrow{k} \mathbb{Z}$ is multiplication by one, the middle map $\mathbb{Z} \xrightarrow{j} \mathbb{Z}$ is multiplication by zero, and the map $\mathbb{Z} \xrightarrow{i=\pi_*} \mathbb{Z}$ is multiplication by one. Hence projection induces the identity map $H_5(E) \xrightarrow{p_*} H_5(S^5)$, and thus the generating 5-cycles in the bundle $E$ are simply related to the generating 5-cycles that wrap the $S^5$ base by projection. A similar argument, using the Wang sequence
$$ \dots \rightarrow H_1(S^4) \rightarrow H_4(S^4) \xrightarrow{i_*} H_4(E) \rightarrow H_0(S^4) \rightarrow H_3(S^4) \rightarrow \dots,$$
where $i$ now denotes the inclusion map $i:S^4\rightarrow E$, tells us that inclusion induces the identity map $H_4(S^4) \xrightarrow{i_*} H_4(E)$, and thus the generating 4-cycles in $E$ are indeed those which wrap the $S^4$ fibre.
}
Correspondingly, we have the non-vanishing de Rham cohomology groups
\begin{equation}
H_{dR}^{4}(E)=H_{dR}^{5}(E)=\mathbb{R}. \label{cohomology}
\end{equation}

Given that the 4th singular homology is non-vanishing,
we must consider worldvolumes whose corresponding 4-cycles are not boundaries. On such non-trivial cycles, we cannot necessarily write a WZ term using Witten's construction, but we can certainly write the action in terms of locally-defined forms in degrees 4, 3, 2, 1, and 0, integrated over chains of the corresponding degree, constructed using \v Cech (co)homology data \cite{Davighi:2018inx}. In fact, we shall soon see that, because of the bundle structure of $(E,\pi)$, a variant of Witten's construction can in fact be carried out, and locally-defined forms (and all the technicalities they entail) will not be needed after all!

\subsection{WZ terms \label{WZ}}

As we have emphasized in \S \ref{review}, there may exist WZ terms corresponding to exact 5-forms. Thus, it is not sufficient to know the cohomology groups (\ref{cohomology});  rather, we need to identify the complete space of $SO(6)$-invariant, integral, closed 5-forms on the target space $E$ that satisfy the Manton condition. Because $G=SO(6)$ is here a semi-simple Lie group, we know from \cite{Davighi:2018inx} that the Manton condition will be automatically satisfied for any $G$-invariant 5-form $\omega$ (even though the Witten construction cannot be used on non-trivial 4-cycles). So, our problem is reduced to finding the space of $SO(6)$-invariant, closed 5-forms on $SO(6)/SO(4)$.

This task in fact reduces to algebra. This is because, given only connectedness of the subgroup $H$, the ring of $G$-invariant forms on $G/H$, which forms a cochain complex under the exterior derivative, is isomorphic to a cochain complex defined algebraically, namely that of the Lie algebra cohomology of $\mathfrak{g}$ relative to $\mathfrak{h}$. This is the space of 
totally antisymmetric maps from $\mathfrak{g}$ to $\mathbb{R}$, which are vanishing on $\mathfrak{h}\subset\mathfrak{g}$ and are $ad\ \mathfrak{h}$-invariant, acted upon by the Lie algebra coboundary operator. We refer the reader to \cite{0031.24803} for the details of this standard construction.

In order to perform this algebraic calculation, and map the resulting space of relative Lie algebra 5-cocycles into a space of WZ terms, we need to introduce local coordinates parametrizing the coset space $SO(6)/SO(4)$. We parametrize the $SO(6)/SO(4)$ cosets by the matrix $U(x)=\exp(\phi_a(x)\hat{T}^a):\Sigma^4\rightarrow SO(6)/SO(4)$, identified up to right multiplication by $H=SO(4)$, where $x$ are the coordinates on the worldvolume $\Sigma^4$, $\{\hat{T}^a\}$ are a basis for the broken generators, and the fields $\phi_a(x)$ define the sigma model map into the target space. 

We choose to embed the $H=SO(4)$ subgroup as the top left 4-by-4 block in $SO(6)$. The nine pNGB fields $\phi_a(x)$ divide into two composite Higgs doublets transforming in the $(\mathbf{2},\mathbf{2})$ of the unbroken $SO(4)\sim SU(2)_L \times SU(2)_R$ subgroup, which we denote by $H_A = (h_A^1,  h_A^2, h_A^3, h_A^4)$ and $H_B = (h_B^1,  h_B^2, h_B^3, h_B^4)$, together with a singlet $\eta$. They are embedded in $\mathfrak{so}(6)$ as follows
\begin{equation}
\phi_a\hat{T}^a = 
\left(
\begin{array}{ccc}
\mathbf{0}_{4\times 4} & H_A^T & H_B^T \\
-H_A & 0 & \eta \\
-H_B & -\eta & 0
\end{array}
\right).
\end{equation}
In our geometric picture, $H_A$ provide local coordinates on the $S^4$ fibre, and the five coordinates $(H_B,\eta)$ provide local coordinates on the $S^5$ base.

Given a suitable basis for the Lie algebras of $SO(6)$ and the $SO(4)$ subgroup as embedded above, we compute the space of closed relative Lie algebra cochains of degree 5 using the \texttt{LieAlgebra[Cohomology]} package in {\tt Maple}. Using the canonical map from the relative Lie algebra cochain complex to the ring of $G$-invariant forms on $G/H$, we identify the following basis for the space of $SO(6)$-invariant closed 5-forms on $E$:
\begin{gather}
\{d^4 H_B d\eta,\ d^4 H_A d\eta,\ \epsilon_{ijkl}dh_A^i  dh_B^j  dh_B^k  dh_B^l d\eta,\\
\qquad \quad \epsilon_{ijkl}dh_A^i  dh_A^j  dh_B^k  dh_B^l d\eta,\ \epsilon_{ijkl} dh_A^i  dh_A^j  dh_A^k  dh_B^l d\eta\},
\end{gather}
where $\epsilon_{ijkl}$ is the usual Levi-Civita symbol with four indices, and we have suppressed the wedges.

We have chosen this basis such that the first element, $d^4 H_B d\eta$, is closed but not exact, and is therefore a representative of the non-trivial 5th cohomology class (\ref{cohomology}), while the remaining four elements are all exact. Given this choice, the first element corresponds to a WZ term with an integer-quantized coefficient, while the others yield real-valued WZ terms. The space of WZ terms in this theory is therefore 
$\mathbb{Z}\times \mathbb{R}^4$.
Note that our chosen representative of the non-trivial cohomology class is simply the pull-back to the bundle $E$ of the evidently $SO(6)$-invariant volume form on the base $S^5$, as one would expect.

Before we move on to discuss the AB term in this model, we now describe more explicitly how these WZ terms in the action can be written. Firstly, the integer-quantized WZ term is unique in that the corresponding 5-form $d^4 H_B d\eta$ can be written as the pull-back to $E$ of a form on $S^5$. Thus, to evaluate the corresponding WZ term, we can in fact push-forward the worldvolume 4-cycle from the target space $E$ to the base $S^5$, using the bundle projection $\pi$, and evaluate the WZ term by performing an integral in the base space. Moreover, since $H_4(S^5,\mathbb{Z})=0$, the push-forward of any 4-cycle to $S^5$ is in fact the boundary of a 5-chain $B$ in the base. The corresponding WZ term evaluated for 4-cycle $z$ is then given, in local coordinates, by the manifestly $SO(6)$-invariant 5-dimensional integral:
\begin{equation}
S_{WZ}[z]=\frac{n}{V_5}\int_B d^4 H_B\ d\eta, \qquad \partial B = \pi_* z,\quad n\in\mathbb{Z},
\end{equation}
So, for this particular term, there is a sense in which Witten's construction goes through, but only after exploiting the bundle structure of the target space.

The remaining four WZ terms correspond to exact 5-forms on $E$, and hence for each we can find a global 4-form $A$ whose exterior derivative is the corresponding 5-form. These terms can therefore all be written as 4-dimensional integrals of globally defined 4-forms over the 4-cycle $z$, each with a different $\mathbb{R}$-valued coefficient. Thus, again, there is no need to introduce locally-defined forms.

\subsection{AB term \label{AB 6over4}} 

The AB term in the action is the integral of a closed (but necessarily not exact) 4-form over the worldvolume 4-cycle, and only depends on the de Rham cohomology class of that 4-form. The 4th de Rham cohomology of $SO(6)/SO(4)$ is one-dimensional (\ref{cohomology}), so we simply need to find a representative of that class. 

Because $G/H$ is compact and $G$ is connected, the de Rham cohomology is in fact isomorphic to the cohomology of $G$-invariant forms, and as stated above, because $H$ is connected, this is furthermore isomorphic to the Lie algebra cohomology of $\mathfrak{g}$ relative to $\mathfrak{h}$.
Hence, we can find such a representative 4-form for our AB term by performing an algebraic calculation in the relative Lie algebra cochain complex, which we again implement in {\tt Maple}.

Such a representative is given by (again suppressing wedges)
\begin{equation}
d^4 H_A + d^4 H_B + \frac{1}{3}\epsilon_{ijkl}\ dh_A^i \ dh_A^j \ dh_B^k \ dh_B^l. \label{AB term}
\end{equation}
Thus, the AB term in the action is locally given by the integral 
\begin{equation}
S_{AB}[z] = \frac{\theta}{2\pi} \int_z \frac{1}{V_4}\left( d^4 H_A + d^4 H_B + \frac{1}{3}\epsilon_{ijkl}\ dh_A^i \ dh_A^j \ dh_B^k \ dh_B^l \right), \quad \theta\in [0,2\pi). \label{6over4 ab}
\end{equation}
As usual, quotienting by the space of integral cohomology classes results in a $U(1)$-valued coefficient for the AB term. Thus, putting everything together, the total space of topological terms in a Composite Higgs model based on the coset $SO(6)/SO(4)$ is given by 
\begin{equation}
\mathbb{Z}\times \mathbb{R}^4 \times U(1).
\end{equation}

\subsection{Twisted versus trivial bundles}

We conclude this section by contrasting the Composite Higgs model on $SO(6)/SO(4)$, which is a (twisted) $S^4$ fibre bundle over $S^5$, with a Composite Higgs model on the corresponding trivial bundle $S^4\times S^5$, which we may realize as the coset space 
\begin{equation}
\frac{SO(5)}{SO(4)}\times \frac{SO(6)}{SO(5)}.
\end{equation}
Let $H_A = (h_A^1,  h_A^2, h_A^3, h_A^4)$ and $(H_B,\eta) = (h_B^1,  h_B^2, h_B^3, h_B^4,\eta)$ provide local coordinates on the $S^4$ and $S^5$ factors respectively (which is of course locally isomorphic to the coordinates introduced above on a patch of $SO(6)/SO(4)$). The transitive action of $G=SO(5)\times SO(6)$ on this space simply factorizes over the two components.

Clearly, the AB term is now simply the integral of the volume form on the $S^4$ factor, {\em viz.} $S_{AB}[z]=(\theta/2\pi V_4)\int_z d^4 H_A$, which is $SO(5)$-invariant and trivially $SO(6)$-invariant. This is precisely analogous to the AB term in the Minimal Model of \S \ref{minimal}. In contrast, in the more complicated $SO(6)/SO(4)$ model above, the $SO(6)$ acts non-trivially on the $S^4$ fibre, such that the volume form on the fibre is not $G$-invariant on its own. 

For the WZ terms, we require an $SO(5)\times SO(6)$-invariant 5-form on this space. Since, in general, the only $SO(n)$-invariant form (in any positive degree) on an $n$-sphere is the volume form, the only such 5-form must be the volume form on the $S^5$ factor. Hence, there is a single WZ term in this model, with quantized coefficient, corresponding to that 5-form. This is precisely analogous to the WZ term in the $SO(6)/SO(5)$ model considered in \S \ref{6over5}. Again, this is in sharp contrast to the more complicated story for $SO(6)/SO(4)$, in which we found a 4-dimensional space of $\mathbb{R}$-valued WZ terms, corresponding to exact, $SO(6)$-invariant 5-forms on $SO(6)/SO(4)$.

In conclusion, we see that even two composite Higgs models which are locally identical, being products of $S^4$ and $S^5$ locally, nevertheless have completely different spectra of topological terms. The differences arise as a subtle interplay between the differing group actions, together with the way that products are globally twisted as bundles.

\section{Two AB terms in the $SO(6)/SO(4)\times SO(2)$ model \label{6over2times4}}

We now consider a variant of the previous two-Higgs-doublet model, in which the linearly realized subgroup $H\subset SO(6)$ is enlarged from $SO(4)$ to $SO(4)\times SO(2)$. This model contains exactly two Higgs doublets, with no singlet $\eta$. A detailed discussion of this model can be found in \cite{Mrazek:2011iu}. Geometrically, the target space is a Grassmannian, that is, the space of planes in $\mathbb{R}^6$.
The story concerning topological terms is much simpler here than in \S \ref{6over4}, because demanding right-$SO(2)$ invariance restricts the basis of projectable forms significantly.

We find that there are no $SO(6)$-invariant forms on this Grassmanian in any odd degree. In particular, there are no $SO(6)$-invariant 5-forms, and so no WZ terms here.

There are, however, invariant forms in even degrees; indeed, there is a 2-dimensional basis of $SO(6)$-invariant 4-forms. Given there are no invariant forms in degrees 3 or 5, these 4-forms are necessarily both closed and not exact, and hence they span a basis for the AB terms in this model:
\begin{gather}
S_{AB}[z] = \frac{\theta_1}{2\pi} \int \frac{1}{N}\left( d^4 H_A + d^4 H_B + \frac{1}{3}\epsilon_{ijkl}\ dh_A^i \ dh_A^j \ dh_B^k \ dh_B^l \right) \\
+ \frac{\theta_2}{2\pi} \int \frac{1}{M}\sum_{ij} dh_A^i\ dh_A^j\  dh_B^i\  dh_B^j,
\end{gather}
where the sum in the second term is over all six pairs of indices $(i,j)$, and both coefficients $\theta_1,\theta_2\in [0,2\pi)$ are periodic. The coefficients $N$ and $M$ are appropriate normalization factors, chosen such that the 4-forms within the integrals are integral.

\section{The Littlest Higgs \label{littlest}}

For our final example, we consider the little Higgs model with coset $SU(5)/SO(5)$.\footnote{The little Higgs models are a subset of Composite Higgs models which exhibit a natural hierarchy between the Higgs vev and the scale of the symmetry breaking $G\rightarrow H$, with the Higgs mass being hierarchically lighter than the other pNGBs. This is achieved by the mechanism of ``collective symmetry breaking", which causes the Higgs potential to be loop-suppressed. For a review of little Higgs models, see Ref. \cite{Schmaltz:2005ky}.} This is the smallest coset known to give a little Higgs, and is therefore known as the ``Littlest Higgs" model \cite{ArkaniHamed:2002qy}. The presence of topological terms in this model was discussed in Ref. \cite{Hill:2007nz}, and has been mentioned in passing elsewhere ({\em e.g.} in \cite{PhysRevD.80.074009}).
Despite this interest, a classification of all topological terms occurring in this model has not been attempted. Indeed, the authors of \cite{Hill:2007nz} merely assert that there is a WZ term in this model, `related to the non-vanishing homotopy group $\pi_5(SU(5)/O(5))=\mathbb{Z}$'. While we shall find that this is essentially the right result, we note that the occurrence of WZ terms in such a sigma model is in fact due to the non-vanishing of the space of $SU(5)$-invariant, closed 5-forms on $SU(5)/SO(5)$, which is unrelated {\em a priori} to the fifth homotopy group.

As we reviewed in \S \ref{review}, there are potentially two types of topological term in a Composite Higgs model (at least that can be written in terms of differential forms): AB terms and WZ terms. The fact that 
\begin{equation}
H_{dR}^4(SU(5)/SO(5),\mathbb{R})=0
\end{equation}
means that there are no AB terms in this model; but there are certainly WZ terms. WZ terms are in one-to-one correspondence with the space of closed, integral, $SU(5)$-invariant 5-forms on $SU(5)/SO(5)$ (because the Manton condition is guaranteed to be satisfied by virtue of $SU(5)$ being semi-simple). We know from the fact that
\begin{equation}
H_{dR}^5(SU(5)/SO(5),\mathbb{R})=\mathbb{R} \label{SU5dR}
\end{equation}
that there is at least one WZ term, because, given compactness of $G/H$ and connectedness of $H$, the de Rham cohomology is isomorphic to the Lie algebra cohomology of $\mathfrak{su(5)}$ relative to $\mathfrak{so}(5)$, which in turn is isomorphic to the cohomology of $SU(5)$-invariant forms on the coset $SU(5)/SO(5)$ \cite{0031.24803}. However, to deduce that this WZ term is unique (up to normalization), we must show that there are no WZ terms corresponding to (de Rham) exact invariant 5-forms. In other words, we must show that the trivial class in the fifth Lie algebra cohomology is empty.

This is indeed the case, as one may show via an explicit calculation using the \texttt{LieAlgebra[Cohomology]} package in {\tt Maple}. In fact, one finds that there are no invariant, exact forms in any degree.\footnote{It is well-known that there are no two-sided $G$-invariant exact forms on $G/H$ if $G/H$ is a symmetric space, which $SU(5)/SO(5)$ is. However, the differential forms that appear in the Relative Lie algebra cohomology (and which correspond to topological terms in our sigma model) are two-sided invariant only for the subgroup $H\subset G$, and one-sided invariant for all of $G$.} 

Thus, the WZ term is indeed unique. The fact that it belongs to a non-trivial cohomology class (in the de Rham sense) means that the restriction to integral classes results in the coefficient of the WZ term being quantized. The upshot is that the space of topological terms in the Littlest Higgs model are indeed classified by a single integer $n\in \mathbb{Z}$. An explicit expression for the WZW term in this case is given in \cite{Hill:2007nz}.

Was it a coincidence that, in this example, the homotopy-based classification yielded the correct answer? While, as we noted, there is \textit{a priori} no direct link between homotopy and cohomology groups, there is of course an indirect link between the two, proceding (via homology) through the Hurewicz map. Indeed, because $SU(5)/SO(4)$ happens to be 4-connected (which means its first non-vanishing homotopy group is $\pi_5=\mathbb{Z}$), the Hurewicz map $h_*:\pi_5(SU(5)/SO(5))\rightarrow H_5(SU(5)/SO(5))$ is in fact an isomorphism. Hence, the fifth homology group, and its dual in singular cohomology, are both $\mathbb{Z}$, from which we deduce  (\ref{SU5dR}). However, the homotopy can certainly tell us nothing about the existence of invariant 5-forms which are exact; in this case, that final piece of information was supplied by an explicit calculation using Lie algebra cohomology.

\section{Connecting the cosets \label{connections}}

In this final section, we discuss how topological terms  in  different Composite Higgs models can in fact be related to each other under RG flow. Firstly, of course, one needs to know how different Composite Higgs models can themselves be related by RG flow.

The idea here is straightforward: if the global symmetry $G$ (which, recall, is spontaneously broken to $H$) is in fact {\em explicitly} broken (via some small parameter) to a subgroup $G'$, then the Goldstones parametrizing the coset space $G/H$ will no longer all be strictly massless. Rather, a potential will turn on for the Goldstones, which will acquire small masses\footnote{By ``small", we mean that the pNGBs will nevertheless remain light relative to the other composite resonances in the theory.} (thus becoming pNGBs). Only the subgroup $H'=G'\cap H$ will then be linearly realized {\em in vacuo}, yielding exact Goldstone bosons on the reduced coset space $G'/H'$. If we flow down to sufficiently low energies, we will be able to integrate out the pNGBs which acquire masses, and thereby arrive at a deep IR theory describing only the massless degrees of freedom. This theory will be a sigma model on $G'/H'$. This concept was recently introduced in Ref. \cite{Setford:2017csx}, under the name of ``Composite Higgs Models in Disguise''.

We postulate that, under such a flow between Composite Higgs Models, the topological terms in the $G/H$ theory should match onto the topological terms in the eventual $G'/H'$ theory.
We now illustrate this proposal with its most simple incarnation, namely the flow between theories based on the cosets:
\begin{equation}
SO(6)/SO(5) \rightarrow SO(5)/SO(4),
\end{equation}
that is, from a theory of Goldstones living on $S^5$, to a theory of Goldstones living on $S^4$.\footnote{Given the theory on $SO(6)/SO(5)\simeq SU(4)/Sp(4)$ has a UV completion (in the form of an $Sp(2N_c)$ gauge theory with an $SU(4)$ flavour symmetry), this provides a model for a UV completion of the Minimal Composite Higgs model, in the form of an $Sp(2N_c)$ gauge theory with an {\em approximate} $SU(4)$ flavour symmetry.} This flow was discussed in \cite{Setford:2017csx}, but we reformulate it here from a more geometric perspective, since this is better suited to a discussion of the topological terms.

\subsection{From the 5-sphere to the 4-sphere}

We begin by considering the sigma model on target space $M=S^5$, which has a transitive group action by $G=SO(6)$. A particular subgroup $G'=SO(5)$ is defined unambiguously by explicit symmetry breaking, as follows. Pick a point $p$ on $M$, which we will define to be the origin in local coordinates $(x_1,\dots, x_5)$. The stabilizer of this point $p$ under the action of $G$ is a subgroup of $G$ isomorphic to $SO(5)$. Define this group to be $G'$, the subgroup of $G$ that remains an exact symmetry of the lagrangian after the explicit breaking is introduced.\footnote{To see how this explicit breaking might be achieved at the level of the lagrangian, we refer the reader to Ref. \cite{Setford:2017csx}.}

Because there is explicit breaking of $SO(6)$, a potential is turned on for the coordinates. What form does it take? We claim that, in suitable coordinates, the potential must be a function of $r^2:=\sum_{i=1}^5 x_i^2$. The reasoning is as follows. The potential $V(x_i)$ must be invariant under the action of the exact symmetry $G'=SO(5)$, which implies that $V(x_i)$ must be constant on the orbits of the $G'$ action. We shall now show that these orbits are indeed surfaces of constant $r$.

Consider an arbitrary point $x_i$ away from the origin. The stabilizer of that point under the original action of $G=SO(6)$ on $M$ is again an $SO(5)$ subgroup of $G$, that is conjugate to $G'$; call this subgroup $H_x$.
The action of the exact symmetry $G'=SO(5)$ at that point $x_i$ is not trivial, so long as $H_x \neq G'$; but there is nevertheless a stabilizer of this $G'$ action given by the intersection of $G'$ with $H_x$. This intersection is an $SO(4)$. So the action of $G'=SO(5)$ traces out orbits which are, by the orbit-stabilizer theorem, isomorphic to
\begin{equation}
G'/(G'\cap H_x) = \left\{
\begin{aligned}
& SO(5)/SO(4)\simeq S^4, \qquad x_i \notin \{0,\bar{0}\}, \\
& SO(5)/SO(5)\simeq \{0\}, \qquad x_i \in \{0,\bar{0}\},
\end{aligned}
\right\}
\label{orbits}
\end{equation}
where $\bar{0}$ denotes the antipodal point on $S^5$ to the origin $0$ (both the origin and its antipode are stabilized by the same subgroup, equal to $G'$; in this sense, the $G'$ action picks out a special pair of points $\{0,\bar{0}\}$). 
Note that, because the $G'$ action on $M$ is not transitive, there need not be only one orbit; in this case, the origin and its antipode are special points, for which the orbit trivially contains only the point itself.
Because the theory is $G'$-invariant, the potential should be constant on each $SO(5)/SO(4)$ orbit through any given non-zero point.

Continuing, if the minimum of the potential is at the origin or its antipode (which are special points with respect to the $G'$ action), we find that there are no massless degrees of freedom (unless the potential equals zero, which just means there is no explicit breaking). But for a minimum at any point which is not the origin, we know from (\ref{orbits}) that there is a whole 4-manifold of degenerate vacua with constant $\sum_{i=1}^5 x_i^2 = a^2 \neq 0$. Thus, there are precisely four Goldstones everywhere (except at the pair of special points), and one massive mode.

Integrating out the massive mode just corresponds (at least at leading order) to restricting to the level set of the minimum of the potential. For the minimum being at the origin, that level set is a point, while for a minimum away from the origin that level set is a 4-sphere, on which the four light degrees of freedom live. Given this $S^4$ has an action of $G'=SO(5)$ (the non-linearly realised global symmetry) with stabilizer $SO(4)=G'\cap H_x$ (the subgroup that is linearly realised), this theory may be identified with the minimal Composite Higgs model.

Looking at it in this way shows that a more convenient set of coordinates is as follows. Let $r=\sqrt{\sum_{i=1}^5 x_i^2}$ be a radial coordinate measuring the distance from the origin, while $\theta_j$, for $j=1,\dots,4$, are four angular coordinates on the level set $S^4$. In these coordinates, we have that the potential $V(r,\theta_j) = V(r)$. We identify the massive radial mode $r$, which is integrated out, with the $\eta$, and the massless angular coordinates $\theta_j$ with the Composite Higgs.

\subsection{From the WZ term to the AB term}

Now we consider the WZ term. As set out in \S \ref{6over5}, the WZ term in the $SO(6)/SO(5)$ theory is proportional to the volume form on $M=S^5$, integrated over a 5-disk $B$ bounding the 4-cycle $z=\partial B$ which defines the field configuration, which may locally be written $S_{WZ}[\partial B]\propto \int_B dr\ d^4 \theta$ in our new coordinates. On such a local patch, the closed 5-form we have integrated is of course exact, and so locally we can re-write $S_{WZ}[\partial B]\propto \int_{\partial B} r\ d^4 \theta$ (more correctly, we can write the WZ term in this way for any cycle $z$ on which the 4-form $r\ d^4 \theta$ is well-defined).
But what happens when we integrate out the massive degrees of freedom? If the minimum is at the origin or its antipode, then all degrees of freedom are massive, and integrated out, so we are left with no dynamics at all, which is clearly uninteresting. So we assume the minimum in $V(r)$ is at some value $r=a$ away from the origin, in which case integrating out the radial mode has the effect of constraining the field configuration to the level set (which is an $S^4$) through $r=a$.

This can be achieved by taking the original 4-cycle $z$ on $S^5$, and pushing it forward onto this level set (under the obvious map $\pi:S^5\rightarrow S^4:\ (r,\theta_j)\mapsto (a,\theta_j)$).\footnote{In other words, we compose the original sigma model map into $S^5$ with the projection $\pi$ onto the level set of $V(a)$ which minimizes the potential on $S^5$.} The 4-form $r\ d^4 \theta$ is well-defined on this level set, and so the WZ term can be written $S_{WZ}[\pi_* z]\propto a\int_{\pi_* } d^4\theta$, which is nothing but the AB term in the Minimal Composite Higgs model defined on the $S^4$ which minimizes $V$.

We shall conclude this section with a few words on how this theory makes contact with the Standard Model electroweak sector, from the geometric perspective we have developed here.
At this level of description, we have a theory which is fully $G'=SO(5)$ invariant, with light degrees of freedom living on $SO(5)/SO(4)$. But to get to the Standard Model, we need to go further. In particular, we need to gauge a subgroup corresponding to the electroweak interactions, which we'll take to be $K=SO(4)$ for ease of description. $K$ must be a subgroup of $G'$, because the interactions that give $r$ a mass should not break the Standard Model gauge symmetry. The gauging also breaks the $SO(6)$ symmetry and leads to another potential on $M=S^5$. What do we know about this potential? It has level sets which are subsets of the level sets of the original potential (because $K \subset G'$ and because the level sets are just the orbits of $K$), but they are now only orbits of $SO(4)$, generically\footnote{Of course, a non-generic miracle is possible: there may be points at which $G'\cap H_x$ coincides with $K$. The level sets here are points. Again, there are no light degrees of freedom about such singular points, and so they are not interesting for us.} with a stabiliser $SO(3)$ ({\em viz.} the intersection of two $SO(4)$ subgroups, $K$ with $G'\cap H_x$). In other words, the true minima of the theory generically have only a non-linearly realised symmetry $K \simeq SO(4)$, of which a subgroup $SO(3)$ is preserved {\em in vacuo}. So, the true vacuum picture is that there are 3 Goldstone bosons (the longitudinal modes of $W^{\pm}$ and $Z$) with an unbroken gauged $SO(3)$ symmetry, corresponding to custodial symmetry. This is precisely the spectrum that we phenomenologically desire.

\section{Discussion \label{discussion}}

In this paper, we have introduced a systematic approach for the identification of topological terms that may appear in the action for a Composite Higgs model. In this approach, which follows the general classification proposed in \cite{Davighi:2018inx}, the possible topological terms divide into two types: Aharonov-Bohm (AB) terms, which correspond to integrating closed, globally-defined 4-forms, and Wess-Zumino (WZ) terms, which correspond to integrating 4-forms that are not closed, and may be only locally-defined, and which moreover must satisfy a non-trivial condition for $G$-invariance called the {\em Manton condition}.

We have applied our classification to a variety of well-studied Composite Higgs models based on different cosets $G/H$, and found topological terms appearing in every one. To summarize, we find AB terms for cosets $SO(5)/SO(4)$, $SO(5)\times U(1)/SO(4)$, $SO(6)/SO(4)$, and $SO(6)/SO(4)\times SO(2)$. In the last example, the space of AB terms is found to be 2-dimensional. We find WZ terms for cosets $SO(6)/SO(5)$, $SO(6)/SO(4)$, and $SU(5)/SO(5)$. In the case of $SO(6)/SO(4)$, the space of WZ terms is isomorphic to $\mathbb{Z}\times \mathbb{R}^4$.

For any given coset, this classification of topological terms is of course exhaustive only to the extent that the assumptions underlying \cite{Davighi:2018inx} are good ones. While this is by and large the case for, say, a Composite Higgs theory, there is one assumption which one might like to relax; this is the assumption that AB terms correspond to closed $4$-forms which are globally-defined on $G/H$.

If we allow AB terms to be constructed from closed $4$-forms which are only locally-defined on $G/H$, one can construct topological terms corresponding to {\em torsion} elements in $H_4(G/H,\mathbb{Z})$.\footnote{A torsion element of an Abelian group is an element of finite order.} Indeed, one may construct such a torsion term as follows.
Consider the map that sends a 4-cycle $z$ (obtained from a worldvolume $\Sigma^4$) to its homology class, and then to its torsion part. Composing this with any map to $U(1)$ defines the action phase on that cycle for a topological term. 
An illustrative example in lower dimension is provided by quantum mechanics of a rigid body, which is described by a 1-d sigma model into target space $SO(3)$. The torsion subgroup of $H_1(SO(3),\mathbb{Z})$ is isomorphic to $\mathbb{Z}/2\mathbb{Z}$, from which one can define a topological term by assigning a relative phase $e^{i\pi}$ to all worldlines in the non-trivial torsion class. Physically, this makes the rigid body fermionic \cite{Davighi:2018inx}. In the case of Composite Higgs models, there will be torsion terms classified by the torsion subgroup of $H_4(G/H,\mathbb{Z})$.

Going further, if we choose to extend our analysis beyond differential forms, and impose more geometric structure on our worldvolume (for example a spin structure), there may be yet further topological terms \cite{Freed:2006mx}. For example, consider a 3-d sigma model with target space $\mathbb{C}P^1$. The dimension of the worldvolume exceeds that of the target space, so there are certainly no AB or WZ terms. Nonetheless there is a topological term, associated with the Hopf invariant $\pi_3(\mathbb{C}P^1)=\mathbb{Z}$, which cannot be written in terms of locally-defined forms (the ``lagrangian'' for this term can only be given as a non-local expression) \cite{Freed:2017rlk}. In fact, requirements of unitarity and locality have been recently used in \cite{Freed:2017rlk} to show that this topological term is only well-defined for certain discrete choices of its coefficient, from which we learn a general lesson: if we seek to extend our classification of topological terms beyond locally-defined differential forms, we must take care to ensure locality and unitarity.

\acknowledgments

We are grateful to Alex Abbott, Nakarin Lohitsiri, Nicholas Manton, Oscar Randal-Williams, and David Tong for discussions. BG is partially supported by STFC consolidated grant ST/P000681/1 and King's College, Cambridge. JD is supported by The Cambridge Trust and STFC consolidated grant ST/P000681/1.

\appendix
\section{Instantons and the physical effects of AB terms \label{instanton}}

To investigate the physical effects of an AB term in a Composite Higgs theory, we consider the Euclidean path integral $\mathcal{Z}$ for the theory. In the case of the MCHM, whose target space is $G/H=SO(5)/SO(4)\simeq S^4$, the partition function $\mathcal{Z}$ is defined by integrating the action phase over the entire space of maps $\phi:\Sigma^4\rightarrow S^4$.

We begin by considering an action consisting of only the two-derivative kinetic term $S_{\mathrm{kin}}$, obtained from an $SO(5)$-invariant metric on the target, together with the AB term $S_{AB}$.
This action is scale-invariant, and admits instanton solutions (which extremize the classical action) in each topological sector ({\em i.e.} in each homotopy class) labelled by $n\in\mathbb{Z}$.\footnote{Since the target space is here an almost quaternionic manifold, there are instantons in each homotopy class corresponding to so-called ``tri-holomorphic maps'' from $\Sigma^4$ to $S^4$, as introduced in \cite{ANSELMI1994255}.}
One can approximate the Euclidean path integral by decomposing it into a sum over topological sectors, and expanding about the saddle points of the classical action in each sector:
\begin{equation}
\mathcal{Z}=\int [\mathcal{D}\phi] e^{-S_{\mathrm{kin}}+iS_{AB}}=\sum_n e^{-S_n+in\theta}K_n,
\end{equation}
where $S_n$ is the classical kinetic term evaluated on an instanton in sector $n$, and $K_n$ is a functional determinant that results from the Gaussian functional integral over  quantum fluctuations. For any given field configuration, the AB term just counts the degree $n$ of the map into the target space.

The factor $K_n$ involves divergent integrals over collective coordinates which parametrize the instanton solutions. Because the two-derivative action is scale-invariant, there will be a collective coordinate $\rho$ parameterising the size of the instanton. We want to know whether the integral over this coordinate diverges for large or small instantons; in other words, in the infrared or the ultraviolet. On purely dimensional grounds, this integral is of the form
\begin{equation}
J=\int \frac{d\rho}{\rho^5} F(\rho\mu),
\end{equation}
where $\mu$ is the renormalization scale, and $F(\rho\mu)$ is a function to be determined. Since $\mathcal{Z}$ is a physical quantity (recall that $-\log \mathcal{Z}$ is the vacuum energy density), the combination $Je^{-S_n}$ must be independent of the renormalization scale $\mu$.

Now, the instanton action $S_n$ depends on the coupling constant in the Composite Higgs theory, which for the kinetic term alone is simply the scale of global symmetry breaking $f$, which, in four spacetime dimensions, has mass dimension one. Since this is a dimensionful coupling, its dependence on the renormalization scale $\mu$ is dominated by the classical contribution. Thus, if we neglect the quantum correction to the running of $f$, the instanton action is independent of $\mu$. Hence, the function $F(\rho\mu)$, needed to ensure RG-invariance, is simply a constant, and the integral over the collective coordinate is just 
\begin{equation}
\int \frac{d\rho}{\rho^5} \sim \rho^{-4},
\end{equation}
which diverges for small instantons, {\em i.e.} in the ultraviolet.

Of course, since $K_n$ is UV divergent, the above calculation is not reliable. What we expect really happens is that at short distances (where instantons give large contributions), the higher derivative terms in the sigma model action become increasingly important relative to the leading two-derivative kinetic term. When these terms are included in the action, the theory will no longer be scale invariant and the instantons will be stabilised at some finite size. Their size will be of order $\Lambda$, where $\Lambda$ is the cut-off for the effective field theory expansion, because the extra terms in the action just feature extra powers of $\partial/\Lambda$.
Our conclusion from all of this is that instantons have a size of order the UV cut-off.

It might be helpful for the reader to compare this 4-d instanton argument with the more familiar story for the theta term in a 2-d sigma model (such as the $\mathbb{C}P^N$ model, in which the AB term is proportional to the integral of the K\"ahler form on $\mathbb{C}P^N$). In two dimensions, the coupling constant $1/g^2$ that appears in front of the kinetic term is dimensionless, and so its running under RG flow is dominated by the 1-loop beta function. The action for an instanton is proportional to $1/g^2$, and thus $e^{-S_n}$ has power-law dependence on the renormalization scale $\mu$. The upshot is an enhancement of the integral over $\rho$ for large $\rho$ due to this 1-loop running, such that the integral in fact diverges in the infrared, and the AB term consequently modifies the vacuum structure of the theory.

\bibliography{articles_bib}
\bibliographystyle{JHEP}
\end{document}